\documentclass[12pt]{iopart}

\pdfoutput=1 
\expandafter\let\csname equation*\endcsname\relax
\expandafter\let\csname endequation*\endcsname\relax
\usepackage{amsmath}
\usepackage{hyperref}
\hypersetup{colorlinks,linkcolor={blue},citecolor={blue},urlcolor={blue}}
\usepackage[top=3cm, bottom=3cm, right=3cm, left=3cm]{geometry} 
\usepackage{multirow}
\usepackage[compress]{cite}
\usepackage{graphicx}
\usepackage{graphics}

\begin{document}

\title{Enhanced Two-Way Teleportation of Entangled States with Six-Qubit Cluster State}

\author[]{Vedhanayagi R$^1$, Soubhik De$^1$, Basherrudin Mahmud Ahmed A$^2$ {\includegraphics[width=0.8em]{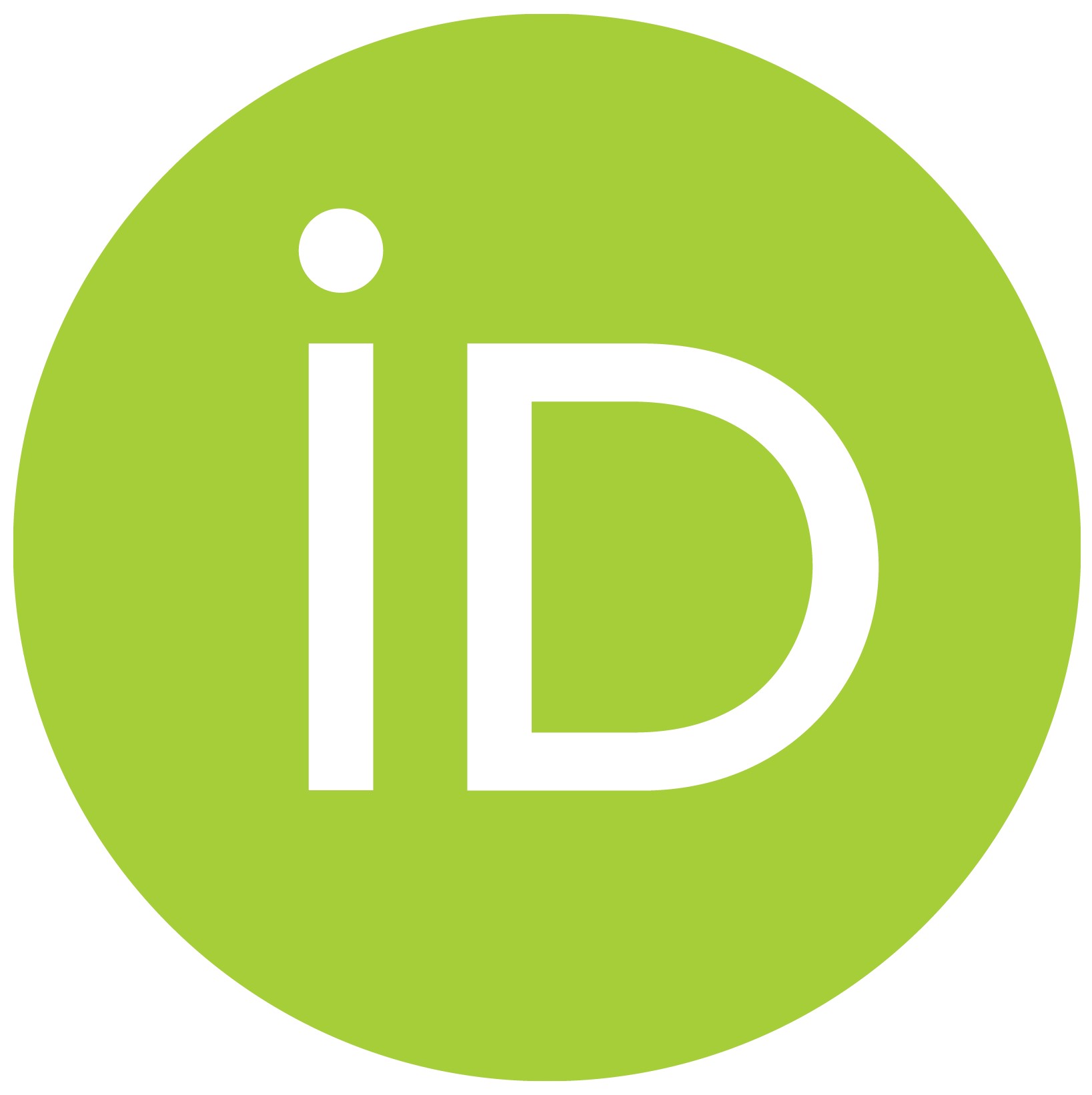}} \href{https://orcid.org/0000-0001-9194-4653} , Alok Sharan$^1$ \href{https://orcid.org/0000-0001-9290-8793}{\includegraphics[width=0.8em]{orcid_1.png}}}

\address{$^1$ Department of Physics, Pondicherry University, Puducherry - 605014, India}
\address{$^2$ Department of Theoretical Physics, School of Physics, Madurai Kamaraj University, Madurai, Tamil Nadu - 625021, India}
\ead{aloksharan@pondiuni.ac.in}
\vspace{10pt}

\begin{abstract}
This work presents a two-way teleportation protocol for the transfer of unknown two-qubit quantum states between two parties, namely Alice and Bob, utilizing a six-qubit cluster state. This bidirectional exchange is achieved by performing Bell measurements on the qubit pairs of Alice and Bob, ensuring the successful teleportation of the quantum state for both parties.  We demonstrate the proposed protocol by designing a teleportation circuit that incorporates the necessary quantum gates. The fidelity of the teleportation process is evaluated through simulator runs, confirming the accuracy and reliability of the proposed scheme. The protocol efficiently reconstructs two-qubit teleported states without requiring additional CNOT operations or auxiliary qubits at the receiver's end, thereby enhancing resource efficiency. A comparative analysis of the intrinsic efficiency with previous approaches establishes that the proposed protocol brings forth an efficient approach for achieving two-way quantum teleportation.
\end{abstract}

\vspace{2pc}

\section{Introduction}
Quantum teleportation is a protocol in quantum communication which facilitates the transfer of quantum states through different parties that are far apart without physically transporting the particles. This process utilizes shared entanglement and classical communication to transfer the quantum information of the system from one party to another. The first proposal for such a teleportation protocol was given by C. H. Bennett et al. in 1993 \cite{Bennett1993}, which Bouwmeester et al. demonstrated later through experiments \cite{Bouwmeester1997}. 
 
In recent years, quantum teleportation has become a foundational protocol for transferring quantum information between multiple parties. Bidirectional quantum teleportation (BQT), a specialized subclass of quantum teleportation is a standard method for two parties to mutually exchange their quantum states. In BQT, both parties share an entangled resource state, perform local operations and measurements, and classical communication to exchange quantum states. The idea of bidirectional quantum teleportation was initially put forward in the work of Vaidman \cite{Vaidman1994}. When the protocol is implemented under the control of supervisor, it forms another variant of BQT, known as bidirectional controlled quantum teleportation (BCQT) \cite{Deng2005,Verma2021}. This protocol is often asymmetric, since the parties typically transfer different types of quantum states, a variant known as asymmetric bidirectional quantum teleportation (ABQT) \cite{Yuan2023}. When asymmetric transfer of quantum states occurs with the supervision of a controlling party, it is referred to as asymmetric bidirectional controlled quantum teleportation (ABCQT) \cite{Choudhury2023}.

Recent research has focused on employing multi-partite states as quantum channels to explore BQT protocols aimed at secure communication. Multiple types of entangled states are being utilized as quantum channels in a number of BQT schemes. Examples include: Bell state \cite{Kiktenko2016}, GHZ state \cite{Zhou2018}, four-qubit cluster state \cite{Kazemikhah2021}, five-qubit cluster state \cite{Kumar2020}, six-qubit entangled state\cite{Xiao2009,Choudhury2009,Zhang2019}, six-qubit cluster state \cite{Zhao2018} and seven-qubit cluster state \cite{Yang2016,Mahjoory2023}. 

Mishra et al. introduced a protocol for two-way quantum communication, where Alice and Bob exchange the two arbitrary unknown quantum states utilizing a special kind six-qubit entangled state \cite{Mishra2011}. Zha et al. were the first to put forth a method for bidirectional quantum controlled teleportation employing a five-qubit cluster state. In this protocol, Alice(Bob) teleports the unknown quantum state of a particle to Bob(Alice), under the supervision of Charlie \cite{Zha2013}. The quantum channel of six-qubit cluster state were used for deterministic teleportation, information splitting, and remote state preparation of an arbitrary two-qubit state \cite{Paul2011}. A scheme for  bidirectional controlled quantum teleportation of single qubit state was proposed using five-qubit entangled channel \cite{Li2013}. Duan et al. demonstrated that Alice and Bob can successfully transmit their single-qubit quantum states to one another via Charlie's control using six qubit entangled channel \cite{Duan2014}. Chen proposed a bidirectional controlled teleportation of an arbitrary single-qubit state utilizing a genuine six-qubit entangled state \cite{Chen2015}. Thapliyal et al. devised a method for choosing appropriate quantum channels for controlled bidirectional teleportation and other controlled quantum communication tasks \cite{Thapliyal2015}. The six-qubit cluster state was utilized as the quantum channel for the symmetric teleportation of a specific three-qubit entangled state \cite{Malik2023}. A multi-hop network within a bidirectional teleportation scheme employing n-GHZ state quantum channel was presented \cite{Zhang2023}.  It has been proposed that the fidelity of the bidirectionally teleported states was enhanced by employing weak measurements and their corresponding reversal operations on the quantum channel \cite{Seida2022}. Yuan et al. proposed a simplified teleportation scheme for six-qubits, employing a six-qubit entangled state as the quantum channel \cite{Yuan2024}. A scheme for cyclic quantum teleportation of three pairs of unknown two-qubit state among three users were proposed, using two six-qubit cluster states as quantum channels.\cite{Slaoui2024}.

This work presents a novel two-way quantum teleportation protocol designed for the bidirectional transfer of unknown two-qubit states among Alice and Bob, utilizing a six-qubit cluster state as the quantum channel. This protocol facilitates the simultaneous exchange of quantum information, enabling Alice to teleport her quantum state to Bob while Bob simultaneously teleports his quantum state to Alice. In this scheme, both Alice and Bob measure their respective qubit pairs in Bell basis, taking one qubit from the input state and another from the shared quantum channel. Following the measurements, the results are communicated via classical channels, allowing each party to apply the necessary unitary operations, thereby faithfully teleporting  the quantum states in both directions. The proposed protocol reconstructs the teleported states without requiring CNOT operations or auxiliary qubits, providing a significant advantage in resource efficiency while preserving fidelity. Further, comparative analysis of its intrinsic efficiency with previous approaches emphasize its superiority.

The structure of the paper is organized as follows: Section 2 provides a theoretical description of the proposed two-way teleportation protocol for two-qubit states using a six-qubit cluster state as the quantum channel. Section 3 outlines the simulation methodology and presents the fidelity analysis results obtained through the IBM quantum computing interface. In Section 4, the success probability of the protocol is calculated, and its intrinsic efficiency is compared with existing protocols. Finally, Section 5 concludes the paper with a summary of the findings.

\section{Two-way Teleportation of Two-Qubit States Through six-qubit Cluster Channel}
A two-way teleportation protocol is presented in this work for the unknown two-qubit states, where Alice can transfer her quantum state $|A\rangle_{A_0 A_1}$ to Bob and Bob can simultaneously teleport his unknown two-qubit state $|B\rangle_{B_0 B_1}$ to Alice by shared six-qubit cluster state quantum channel $|\psi_6\rangle$.

Let us consider that the two participants Alice and Bob each possess unknown two qubit states, given by
\begin{align}
    |A\rangle_{A_0A_1} &=  \big( a_0 |00\rangle + a_1 |11\rangle \big)_{A_0A_1}, \label{eqn1} \\
    |B\rangle_{B_0B_1} &=  \big(b_0 |00\rangle + b_1 |11\rangle \big)_{B_0B_1}, \label{eqn2}
\end{align}  
where the coefficients $a_0, a_1, b_0, b_1$ are complex numbers and the states are normalized such that $|a_0|^2 + |a_1|^2 =1$ and  $|b_0|^2 + |b_1|^2 =1$.

To implement the BQT protocol, Alice and Bob utilize a shared six-qubit cluster state as the quantum channel. The six-qubit cluster state is expressed as follows:
\begin{equation}
    |\psi_6\rangle = \frac{1}{2} \big( |000000\rangle +|101010\rangle +|010101\rangle + |111111\rangle \big)_{123456}.
\end{equation}

\begin{figure}[h]
    \centering
    \includegraphics[width=0.3\linewidth]{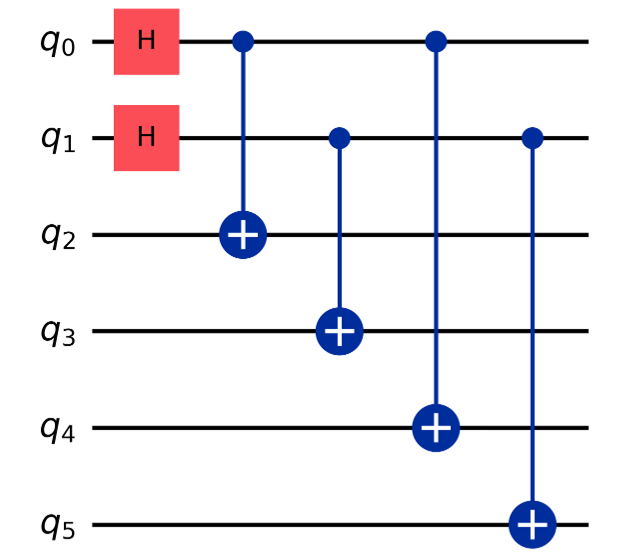}
    \caption{Circuit for six-qubit channel construction}
    \label{fig 1}
\end{figure}

\begin{figure}[h]
    \centering
    \includegraphics[width=0.85\linewidth]{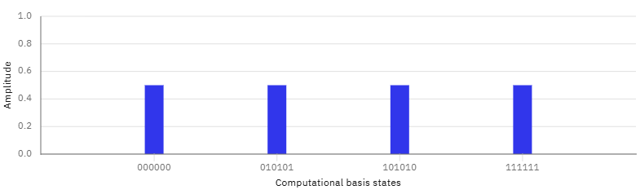}
    \caption{Histogram depicting the amplitudes of the basis states of the six-qubit cluster state.}
    \label{fig 2}
\end{figure}

Alice possesses the qubits $1,4,6$, while qubits $2,3,5$ belong to Bob. The circuit for generating the six-qubit quantum channel, along with its representation in terms of state vector amplitudes corresponding to the computational basis states, with zero amplitude states hidden, are given in Figure \ref{fig 1} and Figure \ref{fig 2} respectively.


In the first step of the teleportation protocol, Alice and Bob each apply a controlled-not (CNOT) gate operation with qubit $A_0$ as control, qubit $A_1$ as target and qubit $B_0$ as control, $B_1$ as target qubit on their respective two qubit input states. The initial states of $|A\rangle_{A_0 A_1} $ and $|B\rangle_{B_0 B_1}$ become
\begin{align}
     |A\rangle_{A_0 A_1} & = ( a_0 |0\rangle + a_1 |1\rangle)_{A_0} \otimes |0\rangle_{A_1}.\\
    |B\rangle_{B_0 B_1} &= (b_0 |0\rangle + b_1 |1\rangle \big)_{B_0} \otimes |0\rangle_{B_1}.
\end{align}    
 
After the CNOT operations, both qubits $A_1$ and $B_1$ will be in the state $|0\rangle$, hence they can be treated as ancillary qubits and be omitted henceforth.
Hence the total state of the system before the teleportation protocol is described by the state
\begin{align}
|\Phi\rangle_{A_0 B_0 123456} &= |A\rangle_{A_0} \otimes |B\rangle_{B_0} \otimes  |\psi_6\rangle, \\
&= ( a_0 |0\rangle + a_1 |1\rangle )_{A_0} \otimes 
   ( b_0 |0\rangle + b_1 |1\rangle )_{B_0}\otimes  \nonumber\\
& \quad \quad \frac{1}{2} \big( |000000\rangle +|101010\rangle +|010101\rangle + |111111\rangle \big)_{123456},\\
&=  \frac{1}{2} \big[a_0b_0(|00000000\rangle + |00101010\rangle + |00010101\rangle +|00111111\rangle)  \nonumber\\ 
& + a_0b_1(|01000000\rangle + |01101010\rangle + |01010101\rangle +|01111111\rangle)  \nonumber\\ 
& +a_1b_0(|10000000\rangle + |10101010\rangle + |10010101\rangle +|10111111\rangle)  \nonumber\\ 
& +a_1b_1(|11000000\rangle + |11101010\rangle + |11010101\rangle +|11111111\rangle)   \big]_{A_0 B_0 
123456}.
\end{align}

To achieve the bidirectional teleportation protocol presented in this work, Alice and Bob conduct Bell measurements on their respective qubit pairs and share their measurement results using classical bits. The Bell basis used for their two-qubit measurements is defined as follows,
\begin{align}
    |\phi^\pm\rangle &= \frac{1}{\sqrt{2}} {|00\rangle \pm |11\rangle},\nonumber \\
    |\psi^\pm\rangle &=  \frac{1}{\sqrt{2}} {|01\rangle \pm |10\rangle}. 
\end{align}    

In the context of quantum circuits, the Bell measurement can be decomposed into CNOT and Hadamard gate operations. When their respective Bell measurements are performed, Alice applies a CNOT gate with $A_0$ as control qubit and $1$ as target qubit followed by a Hadamard gate on $A_0$, and similarly Bob applies a CNOT gate with $B_0$ as control qubit and $2$ as target qubit followed by a Hadamard gate on $B_0$.
\begin{equation}
    |\Phi^\prime\rangle = [CNOT_{A_0,1}\otimes CNOT_{B_0,2} \otimes H_{A_0}\otimes H_{B_0} \otimes I_{3}\otimes I_{4}\otimes I_{5}\otimes I_{6}].
\end{equation}

After the Bell measurements by the two participants, the total state of the quantum system collapses into one of 16 possible states with equal probability, which is given as

\begin{align}
    |\Phi^\prime\rangle  = & \frac{1}{2\sqrt{2}}  \{|\phi^+\rangle_{A_01} |\phi^+\rangle_{B_02} \otimes (a_0 b_0 |0000\rangle + a_0 b_1 |0011\rangle + a_1 b_0 |1100\rangle +a_1 b_1 |1111\rangle )_{3546} \nonumber \\
    + & |\phi^+\rangle_{A_01} |\phi^-\rangle_{B_02} \otimes (a_0 b_0 |0000\rangle + a_0 b_1 |0011\rangle + a_1 b_0 |1100\rangle +a_1 b_1 |1111\rangle )_{3546} \nonumber \\
    + & |\phi^+\rangle_{A_01} |\psi^+\rangle_{B_02} \otimes (a_0 b_0 |0011\rangle + a_0 b_1 |0000\rangle + a_1 b_0 |1111\rangle +a_1 b_1 |1100\rangle )_{3546} \nonumber \\
    + & |\phi^+\rangle_{A_01} |\psi^-\rangle_{B_02} \otimes (a_0 b_0 |0011\rangle - a_0 b_1 |0000\rangle + a_1 b_0 |1111\rangle - a_1 b_1 |1100\rangle )_{3546} \nonumber \\
   + & |\phi^-\rangle_{A_01} |\phi^+\rangle_{B_02} \otimes (a_0 b_0 |0000\rangle + a_0 b_1 |0011\rangle - a_1 b_0 |1100\rangle - a_1 b_1 |1111\rangle )_{3546} \nonumber \\
   + & |\phi^-\rangle_{A_01} |\phi^-\rangle_{B_02} \otimes (a_0 b_0 |0000\rangle - a_0 b_1 |0011\rangle - a_1 b_0 |1100\rangle + a_1 b_1 |1111\rangle )_{3546} \nonumber \\
   + & |\phi^-\rangle_{A_01} |\psi^+\rangle_{B_02} \otimes (a_0 b_0 |0011\rangle + a_0 b_1 |0000\rangle - a_1 b_0 |1111\rangle - a_1 b_1 |1100\rangle )_{3546} \nonumber \\
    + & |\phi^-\rangle_{A_01} |\psi^-\rangle_{B_02} \otimes (a_0 b_0 |0011\rangle - a_0 b_1 |0000\rangle - a_1 b_0 |1111\rangle + a_1 b_1 |1100\rangle)_{3546} \nonumber \\
    + & |\psi^+\rangle_{A_01} |\phi^+\rangle_{B_02} \otimes (a_0 b_0 |1100\rangle + a_0 b_1 |1111\rangle + a_1 b_0 |0000\rangle +a_1 b_1 |0011\rangle )_{3546} \nonumber \\
    + & |\psi^+\rangle_{A_01} |\phi^-\rangle_{B_02} \otimes (a_0 b_0 |1100\rangle - a_0 b_1 |1111\rangle + a_1 b_0 |0000\rangle - a_1 b_1 |0011\rangle )_{3546} \nonumber \\
    + & |\psi^+\rangle_{A_01} |\psi^+\rangle_{B_02} \otimes (a_0 b_0 |1111\rangle + a_0 b_1 |1100\rangle + a_1 b_0 |0011\rangle +a_1 b_1 |0000\rangle )_{3546} \nonumber \\
    + & |\psi^+\rangle_{A_01} |\psi^-\rangle_{B_02} \otimes (a_0 b_0 |1111\rangle - a_0 b_1 |1100\rangle + a_1 b_0 |0011\rangle - a_1 b_1 |0000\rangle )_{3546} \nonumber \\
    + & |\psi^-\rangle_{A_01} |\phi^+\rangle_{B_02} \otimes (a_0 b_0 |1100\rangle + a_0 b_1 |1111\rangle - a_1 b_0 |0000\rangle -a_1 b_1 |0011\rangle )_{3546} \nonumber \\
    + & |\psi^-\rangle_{A_01} |\phi^-\rangle_{B_02} \otimes (a_0 b_0 |1100\rangle - a_0 b_1 |1111\rangle - a_1 b_0 |0000\rangle +a_1 b_1 |0011\rangle )_{3546} \nonumber \\
    + & |\psi^-\rangle_{A_01} |\psi^+\rangle_{B_02} \otimes (a_0 b_0 |1111\rangle + a_0 b_1 |1100\rangle - a_1 b_0 |0011\rangle - a_1 b_1 |0000\rangle )_{3546} \nonumber \\
    + & |\psi^-\rangle_{A_01} |\psi^-\rangle_{B_02} \otimes (a_0 b_0 |1111\rangle - a_0 b_1 |1100\rangle + a_1 b_0 |0011\rangle - a_1 b_1 |0000\rangle )_{3546} \}
\end{align}

For example: After Alice and Bob's measurements, one of the possible collapsed state that resulted from the above equation is
\begin{equation}
    |\Phi' \rangle = |\psi^+\rangle_{A_01} \otimes |\phi^+\rangle_{B_02} \otimes (a_0 b_0 |1100\rangle + a_0 b_1 |1111\rangle + a_1 b_0 |0000\rangle +a_1 b_1 |0011\rangle )_{3546}.  
\end{equation}

This state can be rearranged such that $a_0$ and $a_1$ coefficients are present in qubits $3$ and $5$, and $b_0$ and $b_1$ are present in qubits $4$ and $6$, as given in Eqn \ref{eqn13}. Now, Alice applies the unitary operator $I_4 \otimes I_6$ to her qubits $4$ and $6$, transforming her qubit state into the form of Bob's initial unknown state $|B\rangle$, as given in Eqn \ref{eqn2}. Similarly, Bob perform the unitary operator $X_3 \otimes X_5$ to his qubit state $3$ and $5$ thereby transforming his qubit state into the form of Alice's initial unknown state $|A\rangle$, as defined in Eqn \ref{eqn1}.
\begin{equation}
    \label{eqn13}
       |A\rangle_{46} \otimes |B\rangle_{35} =(b_0|00\rangle+b_1|11\rangle)_{46}  \otimes (a_0|11\rangle+ a_1|00\rangle)_{35}. 
\end{equation}

\begin{figure}
    \centering
    \includegraphics[width=1.0\linewidth]{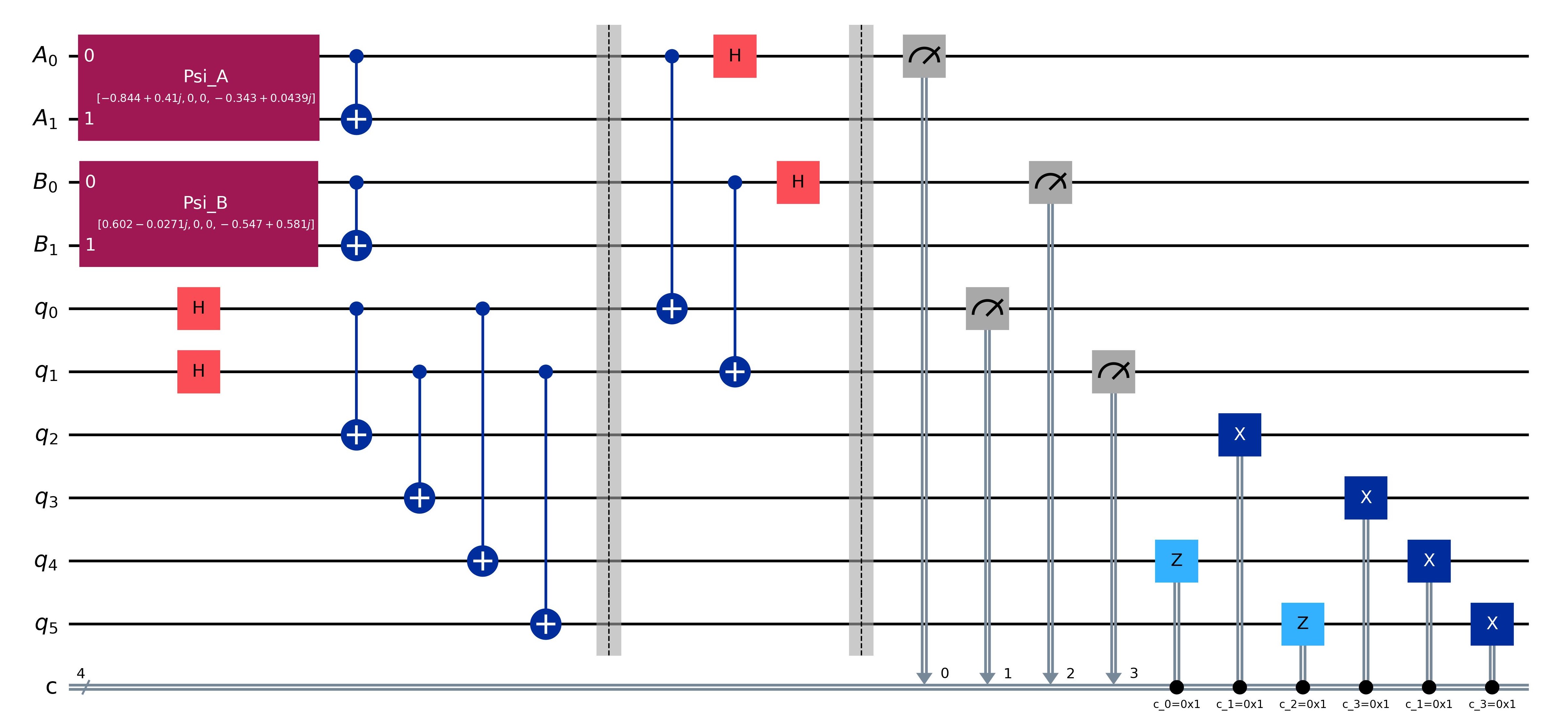}
    \caption{Circuit illustrating the two-way teleportation of two qubit states through six-qubit cluster channel. Alice's and Bob's input states are stored in qubits $A_0$ and $A_1$, and $B_0$ and $B_1$, respectively. The qubits $q_0,q_1,...q_5$ represents the six-qubit quantum channel}
    \label{fig 3}
\end{figure}

Similarly for all other possible outcomes, Alice implements the suitable unitary operations on her qubits $4$ and $6$ to retrieve the quantum information from Bob's input state $|B\rangle$. Likewise, Bob applies the appropriate unitary operations to his qubits $3$ and $5$ to reconstruct the quantum information encoded in Alice's input state $|A\rangle$. All possible measurement outcomes for Alice's and Bob's qubits, along with their corresponding unitary rotations, are summarized in Table \ref{tab 1}.
\begin{table}[h]
    \centering
    \caption{The possible measurement results of Alice and Bob states and their associated unitary rotations (UR)}
     \label{tab 1}
      \setlength{\tabcolsep}{4pt}
     \renewcommand{\arraystretch}{1.2}
     \small
     \begin{tabular}{@{}lllll}
    \br
    Alice's state & Bob's state & Collapsed state & Bob UR & Alice UR \\
         \mr
        $|\phi^+\rangle_{A_01}$ & $|\phi^+\rangle_{B_02}$ & $ (a_0|00\rangle+a_1|11\rangle)_{35} \otimes (b_0|00\rangle+b_1|11\rangle)_{46}$ & $I_{3} \otimes I_{5}$ & $I_{4} \otimes I_{6}$  \\
        $|\phi^+\rangle_{A_01}$ & $|\phi^-\rangle_{B_02}$ & $ (a_0|00\rangle+a_1|11\rangle)_{35} \otimes (b_0|00\rangle-b_1|11\rangle)_{46}$ & $I_{3} \otimes I_{5}$ & $I_{4} \otimes Z_{6}$  \\ 
        $|\phi^+\rangle_{A_01}$ & $|\psi^+\rangle_{B_02}$ & $ (a_0|00\rangle+a_1|11\rangle)_{35} \otimes (b_0|11\rangle+b_1|00\rangle)_{46}$ &  $I_{3} \otimes I_{5}$ & $X_{4} \otimes X_{6}$  \\ 
        $|\phi^+\rangle_{A_01}$ & $|\psi^-\rangle_{B_02}$ & $ (a_0|00\rangle+a_1|11\rangle)_{35} \otimes (b_0|11\rangle-b_1|00\rangle)_{46}$ & $I_{3} \otimes I_{5}$ & $Z_4X_{4} \otimes X_{6}$  \\ 
        $|\phi^-\rangle_{A_01}$ & $|\phi^+\rangle_{B_02}$ & $ (a_0|00\rangle -a_1|11\rangle)_{35} \otimes (b_0|00\rangle+b_1|11\rangle)_{46}$ &  $I_{3} \otimes Z_{5}$ & $I_{4} \otimes I_{6}$  \\ 
        $|\phi^-\rangle_{A_01}$ & $|\phi^-\rangle_{B_02}$ & $ (a_0|00\rangle -a_1|11\rangle)_{35} \otimes (b_0|00\rangle-b_1|11\rangle)_{46}$ &  $I_{3} \otimes Z_{5}$ & $I_{4} \otimes Z_{6}$  \\ 
        $|\phi^-\rangle_{A_01}$ & $|\psi^+\rangle_{B_02}$ & $ (a_0|00\rangle -a_1|11\rangle)_{35} \otimes (b_0|11\rangle-b_1|00\rangle)_{46}$ &  $I_{3} \otimes Z_{5}$ & $X_{4} \otimes X_{6}$  \\ 
        $|\phi^-\rangle_{A_01}$ & $|\psi^-\rangle_{B_02}$ & $ (a_0|00\rangle -a_1|11\rangle)_{35} \otimes (b_0|11\rangle-b_1|00\rangle)_{46}$ &  $I_{3} \otimes Z_{5}$ & $Z_4X_{4} \otimes X_{6}$  \\ 
        $|\psi^+\rangle_{A_01}$ & $|\phi^+\rangle_{B_02}$ & $ (a_0|11\rangle+a_1|00\rangle)_{35} \otimes (b_0|00\rangle+b_1|11\rangle)_{46}$ &  $X_{3} \otimes X_{5}$ & $I_{4} \otimes I_{6}$  \\ 
        $|\psi^+\rangle_{A_01}$ & $|\phi^-\rangle_{B_02}$ & $ (a_0|11\rangle+a_1|00\rangle)_{35} \otimes (b_0|00\rangle-b_1|11\rangle)_{46}$ & $X_{3} \otimes X_{5}$ & $I_{4} \otimes Z_{6}$  \\ 
        $|\psi^+\rangle_{A_01}$ & $|\psi^+\rangle_{B_02}$ & $ (a_0|11\rangle+a_1|00\rangle)_{35} \otimes (b_0|11\rangle+b_1|00\rangle)_{46}$ & $X_{3} \otimes X_{5}$ & $X_{4} \otimes X_{6}$  \\ 
        $|\psi^+\rangle_{A_01}$ & $|\psi^-\rangle_{B_02}$ & $ (a_0|11\rangle+a_1|00\rangle)_{35} \otimes (b_0|11\rangle-b_1|00\rangle)_{46}$ & $X_{3} \otimes X_{5}$ & $Z_4X_{4} \otimes X_{6}$  \\ 
        $|\psi^-\rangle_{A_01}$ & $|\phi^+\rangle_{B_02}$ & $ (a_0|11\rangle-a_1|00\rangle)_{35} \otimes (b_0|00\rangle+b_1|11\rangle)_{46}$ &  $Z_3X_{3} \otimes X_{5}$ & $I_{4} \otimes I_{6}$  \\
        $|\psi^-\rangle_{A_01}$ & $|\phi^-\rangle_{B_02}$ & $ (a_0|11\rangle-a_1|00\rangle)_{35} \otimes (b_0|00\rangle-b_1|11\rangle)_{46}$ & $Z_3X_{3} \otimes X_{5}$ & $I_{4} \otimes Z_{6}$  \\ 
        $|\psi^-\rangle_{A_01}$ & $|\psi^+\rangle_{B_02}$ & $ (a_0|11\rangle-a_1|00\rangle)_{35} \otimes (b_0|11\rangle+b_1|00\rangle)_{46}$ & $Z_3X_{3} \otimes X_{5}$ & $X_{4} \otimes X_{6}$  \\ 
        $|\psi^-\rangle_{A_01}$ & $|\psi^-\rangle_{B_02}$ & $ (a_0|11\rangle-a_1|00\rangle)_{35} \otimes (b_0|11\rangle-b_1|00\rangle)_{46}$ & $Z_3X_{3} \otimes X_{5}$ & $Z_4X_{4} \otimes X_{6}$  \\ 
        \br
    \end{tabular}    
\end{table}
In this manner, Alice and Bob successfully retrieve each others' 2-qubit states, thereby completing the bidirectional teleportation protocol. A key advantage of the proposed protocol is its ability to recover the original 2-qubit input states for both Alice and Bob at the receiver's end without requiring auxiliary qubits or additional CNOT operations. Notably, the output states preserve the two-qubit structure of the input state prior to CNOT operation, ensuring consistency and resource efficiency. This approach ensures efficient utilization of quantum resources while maintaining the fidelity of the protocol. Thus, two-way quantum teleportation of an unknown two-qubit state has been successfully achieved using a six-qubit cluster state. The circuit depicted in Figure \ref{fig 3} can be used to implement the proposed scheme. 

\section{Results}
This section details the experimental validation of the proposed teleportation protocol, implemented using the IBM Qiskit interface. The message state qubits to be teleported are denoted as $A_0$ and $A_1$ representing Alice's input state $|A\rangle$, and $B_0$ and $B_1$ representing Bob's input state $|B\rangle$. In our quantum circuit, the six-qubit quantum channel is formed by qubits $q_0$ through $q_5$, where qubits $q_0, q_3$ and $q_5$ are assigned to Alice and qubits $q_1, q_2$ and $q_4$ are assigned to Bob. 
To determine the accuracy of our scheme, the fidelity of the teleportation protocol is evaluated. Teleportation fidelity quantifies how closely the teleported quantum state matches with the original input state. Experimentally, teleportation fidelity of the protocol is calculated by preparing a set of arbitrary input states, teleporting them through the quantum circuit and reconstructing the density matrices of the input and output states using the measurement counts of the experiment.

In this study, a set of arbitrary two-qubit states were randomly generated and assigned as Alice's input state $|A\rangle_{A_0 A_1} $ and Bob's input state $|B\rangle_{B_0 B_1}$. The density matrix form of these two input states $\rho_A$ and $\rho_B$ were evaluated, as described by the following equations
\begin{align}
     \rho_{A} &= (|A\rangle \langle A| )_{A_0 A_1} \nonumber\\
    \rho_{B} &= (|B\rangle \langle B|)_{B_0 B_1}
\end{align}

After teleportation, the total quantum state of the system comprises of the teleported qubits $|A\rangle_{A_0 A_1}$ and $|B\rangle_{B_0 B_1}$ along with the remaining qubits $q_0$ to $q_5$ in the quantum channel. The post-teleportation states $|A\rangle_{46}$ and $|B\rangle_{35}$ are obtained by tracing out the other qubits, after which the density matrices of the output states are constructed through the outer product, given in Eqn \ref{eqn15}. The fidelity is calculated by comparing the output state's density matrices $\sigma_A$ and $\sigma_B$ with input state's density matrices $\rho_A$ and $\rho_B$ for the respective participants Alice and Bob.
\begin{align}
\label{eqn15}
     \sigma_{A} &= (|A\rangle \langle A|)_{46} \nonumber\\
    \sigma_{B} &= (|B\rangle \langle B|)_{35}
\end{align}

The fidelity of teleportation $F$ is expressed as,
\begin{equation}
      F(\rho, \sigma) = \bigg (Tr\big( \sqrt{\sqrt{\rho} \cdot \sigma \cdot \sqrt{\rho}}\big)\bigg)^2.
\end{equation}

In bidirectional teleportation, Alice's quantum state information is transferred to Bob and vice versa. As a result, Alice's output state $|A\rangle_{46}$ is equivalent to Bob's input state $|B\rangle_{B_0 B_1}$ and Bob's output state $|B\rangle_{35}$ is equivalent to Alice's input state $|A\rangle_{A_0 A_1}$. Hence, the teleportation fidelities for Alice and Bob, given by $F_A$ and $F_B$, are calculated using the following equations
\begin{align}
    F_A =& \bigg (Tr\big( \sqrt{\sqrt{\rho_A} \cdot \sigma_B \cdot \sqrt{\rho_A}}\big)\bigg)^2. \\
    F_B =& \bigg (Tr\big( \sqrt{\sqrt{\rho_B} \cdot \sigma_A \cdot \sqrt{\rho_B}}\big)\bigg)^2.
\end{align}

\begin{figure}[!ht]
    \centering
    \includegraphics[width=0.6\linewidth]{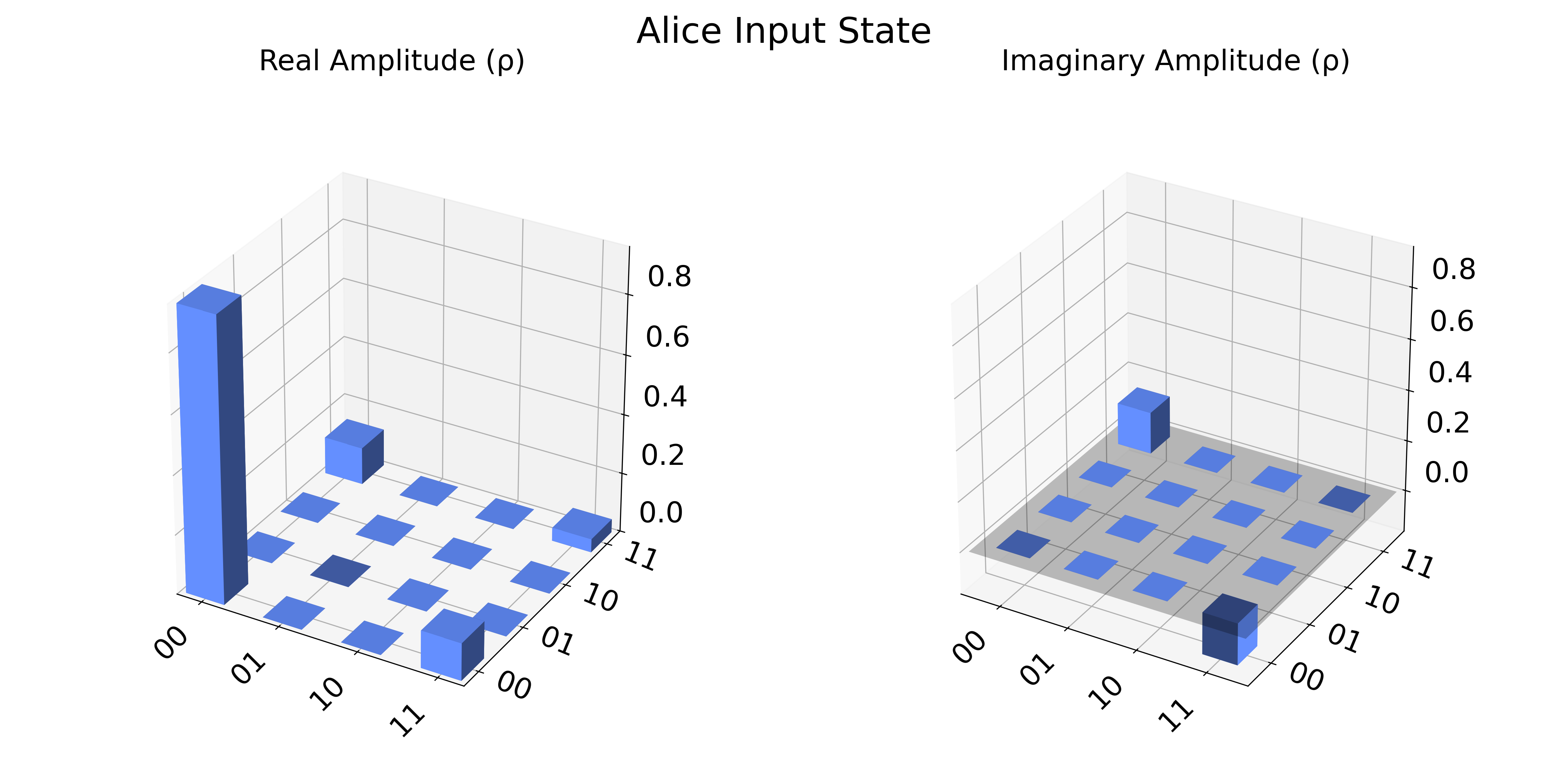}\\
    \vspace{0.2cm} 
    \includegraphics[width=0.6\linewidth]{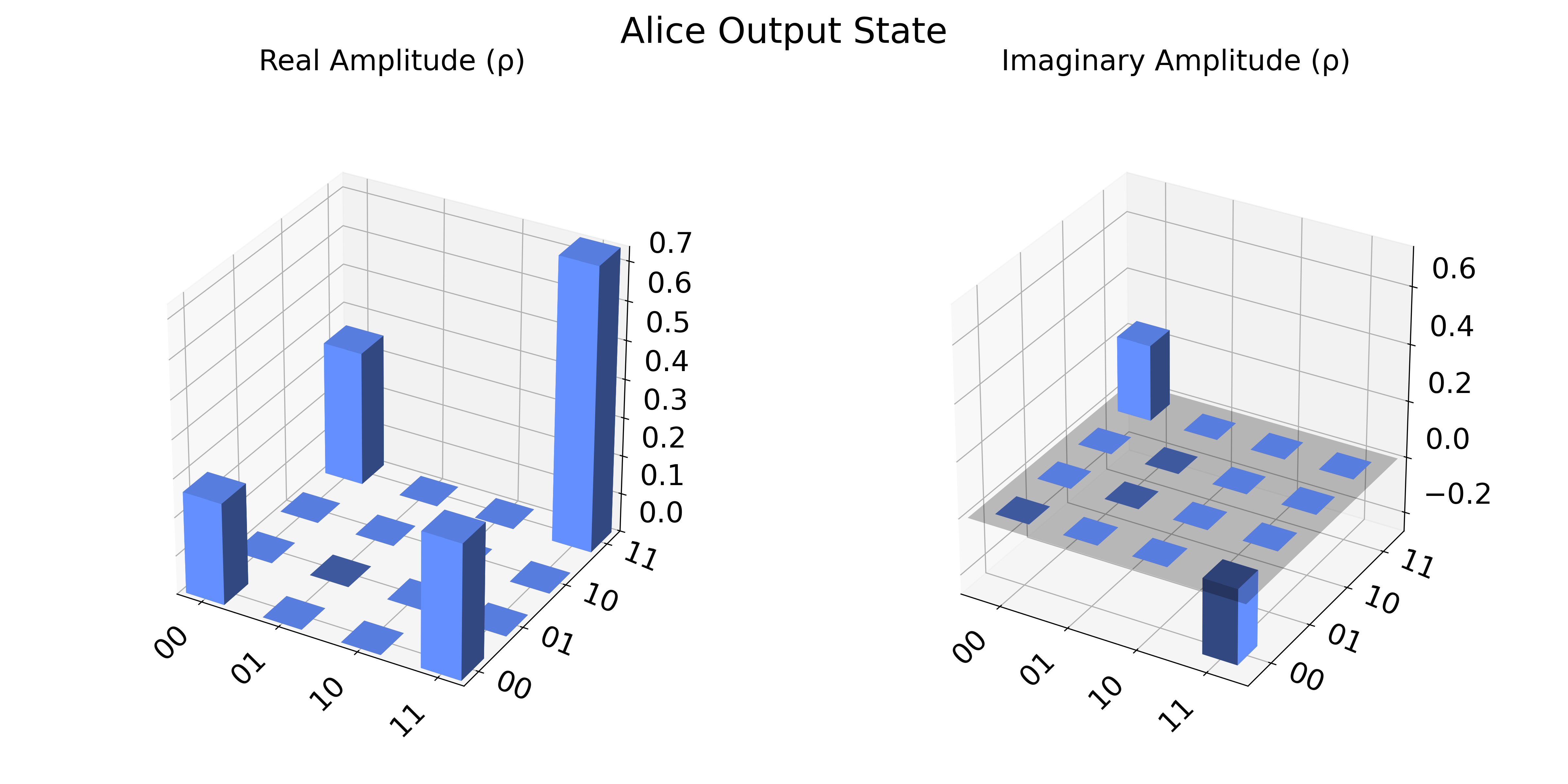}
    \caption{ Real and imaginary parts of Alice's and Bob's input state density matrices }
    \label{fig 4}
\end{figure}

\begin{figure}[!ht]
    \centering
    \includegraphics[width=0.6\linewidth]{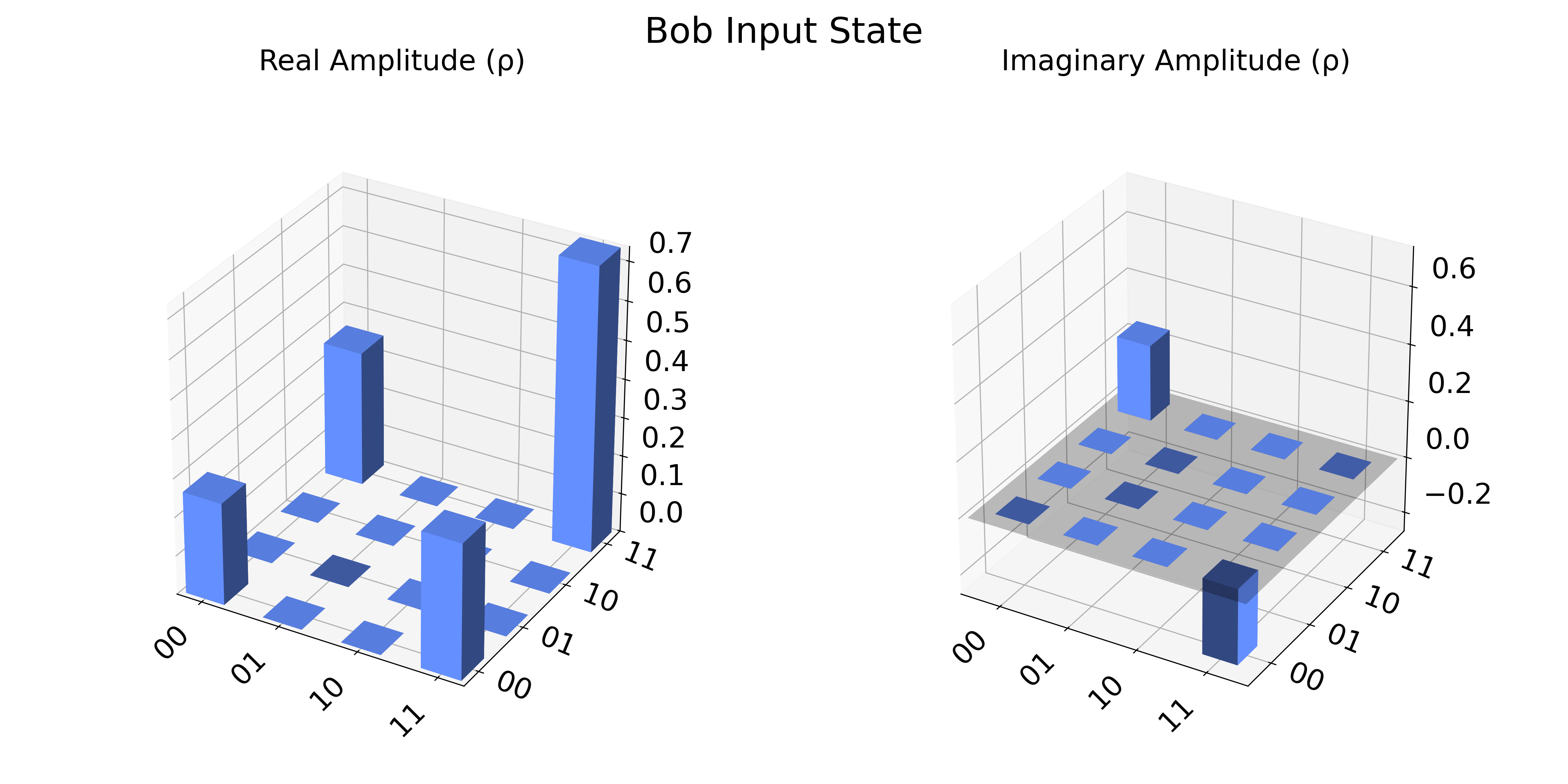}\\
    \vspace{0.2cm} 
    \includegraphics[width=0.6\linewidth]{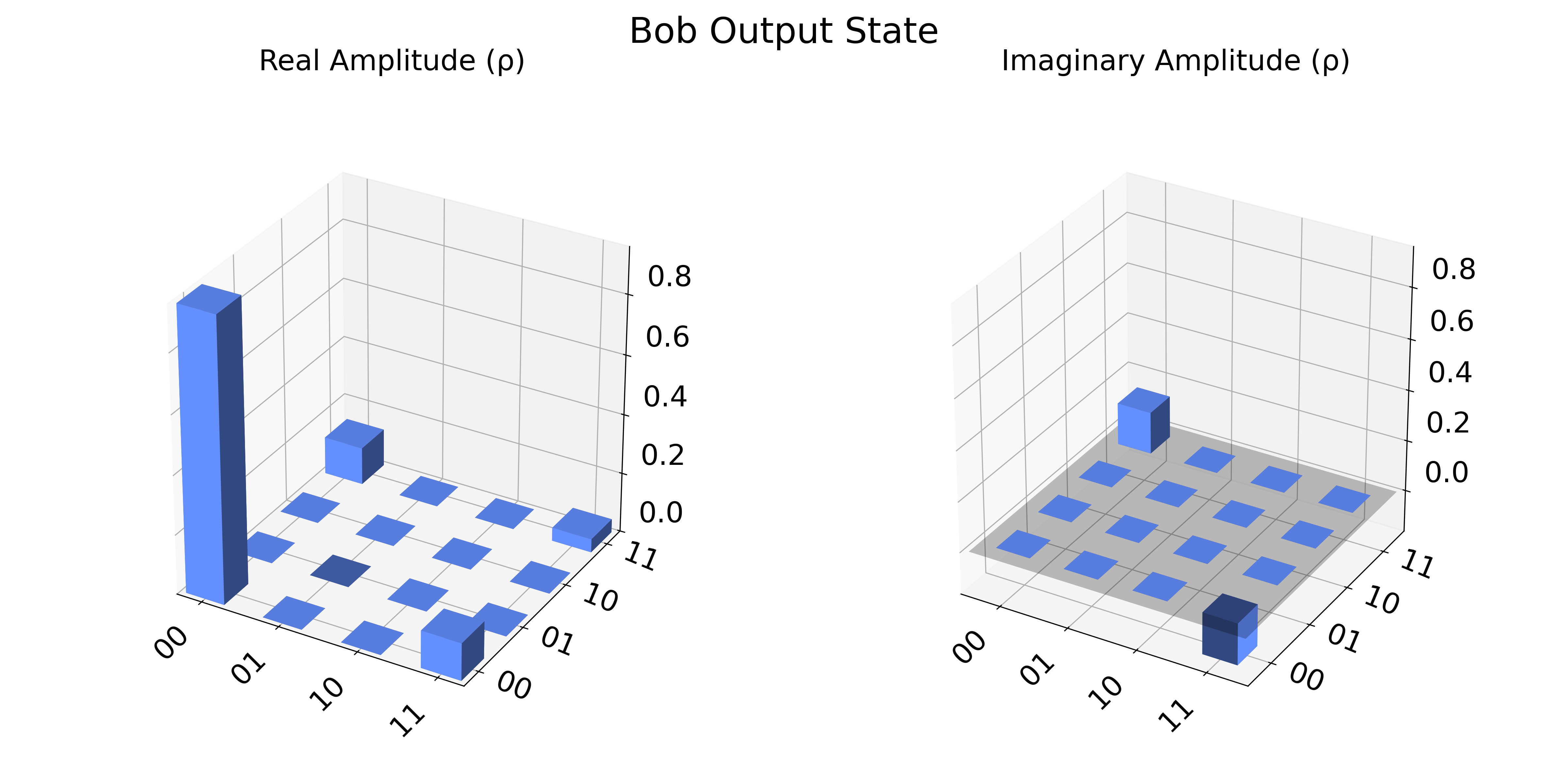}
    \caption{Real and imaginary parts of Alice's and Bob's output state density matrices}
    \label{fig 5}
\end{figure}

The teleportation fidelities of Alice and Bob are both $\approx 1$. Figure \ref{fig 4} presents the state-city plot of the Alice and Bob's input density matrices $\rho_A$ and $\rho_B$, showing their real and imaginary components. Similarly, Figure \ref{fig 5} presents the state-city plot of the Alice and Bob's output density matrices $\sigma_A$ and $\sigma_B$, showing their real and imaginary components. We note that the state-city plot for Alice's input state and Bob's output state are similar, and so are Bob's input state and Alice's output state. This confirms the fact that the calculated fidelities are $\approx 1$.

\section{Discussion}
Let us now analyze the key features of the proposed scheme, focusing on two aspects: success probability and intrinsic efficiency. First, we consider the success probability of the proposed protocol, which is defined as follows \cite{Shi2018},
\begin{align}
    \textit{P} &= \sum_{i=1}^{m} \left(\frac{1}{N}\right)_i, \\
    \implies P &= \sum_{i=1}^{16} \left(\frac{1}{16}\right)_i = 1.
\end{align}

Here, $N$ denotes the total number of possible outcomes for the Alice and Bob measurement, while $m$ indicates the number of measurement outcomes that counts for successful reconstruction of the states \cite{Yuan2024}. In our protocol, all 16 possible measurement outcomes result in the successful reconstruction of the unknown states, i.e, $N$ and $m$ are both 16. Therefore, the overall success probability of the proposed protocol is 1.

Next, we examine the intrinsic efficiency of the teleportation protocol, which refers to its ability to transmit quantum information optimally using available quantum resources. The efficiency $\eta$ of the teleportation scheme is defined in \cite{Yuan2008}
\begin{equation}
    \eta = \frac{q_s}{q_u+q_a+b_t},
\end{equation}

where $q_s$ is the number of message qubits consisting the teleported quantum information, $q_u$ is the number of resource qubits used as the quantum channel, $q_a$ refers to the number of auxiliary qubits and $b_t$ is the number of classical bits transmitted. The efficiency of our protocol is calculated to be $40\%$, indicating that it offers higher transmission efficiency compared to other protocols.

A detailed comparison between the proposed scheme and previously established bidirectional schemes from literature is presented in Table \ref{tab 2}, which illustrates the advantages of the proposed scheme. In the table, BQT represents the type of bidirectional teleportation, QIBT corresponds to the number of qubits being teleported, QR refers to the number of quantum resource qubits, CR represents classical resource bits, AQ stands for number of auxiliary qubits, and $\eta$ is the calculated intrinsic efficiency.
\begin{table}[h]
    \centering 
    \caption{Comparison between different BQT Schemes}
    \label{tab 2}
    \setlength{\tabcolsep}{6pt}
    \renewcommand{\arraystretch}{1}
    \begin{tabular}{@{}l l l l l l l}
    \hline
    Scheme & BQT & QIBT & QR & CR & AQ & Efficiency $(\eta)$ \\
    \hline 
        \cite{Yan2013} & 1 $\leftrightarrow$ 1 & 2 & 6 & 6 & 0 &  16.66$\%$\\
        \cite{Li2016} & 2 $\leftrightarrow$ 1 & 3 & 6 & 6 & 0 & 25 $\%$\\
        \cite{Zhou2019} & 3 $\leftrightarrow$ 1 & 4 & 6 & 6 & 0 & 33.3$\%$ \\
        \cite{Hassanpour2016} & 2 $\leftrightarrow$ 2 & 4 & 6 & 6 & 0 & 33.3 $\%$\\
        \cite{Zhou2020} & 2 $\leftrightarrow$ 2 & 4 & 6 & 6 & 0 & 33.3 $\%$\\
        \cite{Zhou2020}& 2 $\leftrightarrow$ 3 & 5 &  6 & 6 & 1 & 38.4 $\%$\\
        \cite{Yuan2020} & 1 $\leftrightarrow$ 1 & 2 & 6 & 6 & 0 & 16.66$\%$ \\ 
        \cite{Chen2020}& 2(1) $\leftrightarrow$ 2 & 4 & 8 & 8 & 0 & 31.25 $\%$\\
        \cite{Choudhury2021}& 3 $\leftrightarrow$ 2 & 5 & 10 & 18 & 0 & 17.85$\%$ \\
        \cite{Verma2020}& 3 $\leftrightarrow$ 1 & 4 & 6 & 5 & 0 & 36.4$\%$ \\
        \cite{Wang2022} & 2 $\leftrightarrow$ 1 & 3 & 5 & 5 & 0 & 30$\%$ \\
        \cite{Dai2022}& 3 $\leftrightarrow$ 2 & 5 & 7 & 6 & 0 & 38.5$\%$ \\
        \cite{Sisodia2023}& 2 $\leftrightarrow$ 2 & 4 & 4 & 4 & 2 & 40 $\%$ \\ 
        Ours & $ 2\leftrightarrow 2 $ & 4 & 6 & 4 & 0 & $40 \%$ \\
    \hline
    \end{tabular}   
\end{table}

\section{Conclusion}

In this work, we present a novel two-way teleportation protocol that enables the exchange of quantum states between two parties, employing a six-qubit cluster state as the quantum channel. The protocol effectively employs Bell measurements on qubit pairs of Alice and Bob, allowing them to exchange measurement results and facilitate the teleportation process. This allows Alice to faithfully teleport her two-qubit state to Bob, while  enabling the teleportation of Bob's quantum state to Alice. The ability of the proposed protocol to recover the original two-qubit input states for both Alice and Bob without requiring auxiliary qubits or additional CNOT operations at both the receivers' ends provides a key advantage, enabling efficient resource utilization while maintaining fidelity. We demonstrated the implementation and simulation runs of the proposed protocol using quantum computing tools. The simulation results using IBM Qiskit confirm the robustness and accuracy of the protocol, with a fidelity of $\approx 1$. Furthermore, the comparison of the intrinsic efficiency of our scheme with previous works signifies that our protocol achieves a higher transmission efficiency while maintaining the success probability of 1.

The implementation of this protocol with current quantum technologies opens the path for experimental validation and practical applications. It provides a promising foundation for developing advanced quantum communication protocols, with potential uses in distributed quantum networks and quantum computing. Future research could focus on improving scalability and practical implementation in real-world quantum systems. Further, work can also be directed towards incorporating controllers into the protocol and modifying it to enable bidirectional teleportation of general two-qubit states.

\section*{Ethics Statements}
\section*{Acknowledgment}
We gratefully acknowledge the support of the MeitY Quantum Computing Applications Lab (QCAL) in association with AWS (Amazon Web Services) for providing access to the Amazon Braket cloud computing platform. We also extend our sincere gratitude to IBM Qiskit for facility quantum simulations.
\section*{Competing interests}
The authors have no competing interests to declare.

\section*{Funding}
No funds or grants were received during the course of this manuscript.

\section*{ORCID iDs}
Alok Sharan {\includegraphics[width=0.8em]{orcid_1.png} } \href{ https://orcid.org/0000-0001-9290-8793} - https://orcid.org/0000-0001-9290-8793\\
A Basherrudin Mahmud Ahmed {\includegraphics[width=0.8em]{orcid_1.png}} \href{https://orcid.org/0000-0001-9194-4653} - https://orcid.org/0000-0001-9194-4653

\end{document}